\documentclass[sigconf,nonacm]{acmart}

\usepackage{balance}
\usepackage{tabularx}
\usepackage{tikz}
\usetikzlibrary{positioning,arrows.meta,fit,backgrounds}

\IfFileExists{libertine.sty}{}{\usepackage{lmodern}}


\title[Beyond Named Mechanisms]{Beyond Named Mechanisms: Economic Institutions as a Design Space for Artificial Agents}

\author{Nikolos Gurney}
\affiliation{
  \institution{Institute for Creative Technologies, University of Southern California}
  \city{Los Angeles}
  \state{CA}
  \country{United States}}
\email{gurney@ict.usc.edu}

\author{Fred Morstatter}
\affiliation{
  \institution{Information Sciences Institute, University of Southern California}
  \city{Marina del Rey}
  \state{CA}
  \country{United States}}
\email{fred@isi.edu}

\begin{abstract}
Multiagent research inherited its economic vocabulary of auctions, markets, prices, bargaining, and matching from a single observed case: human societies. Those institutions bundle identity, competence, execution, ownership, commitment, and settlement in ways that suit agents whose minds are private, who learn rather than copy skills, and who each persist as one person. Engineered agents need not share any of these constraints, making the list of familiar institutions look less like the entirety of economic organization and more like a sample from a larger, unmapped space. We sketch one way to make that space concrete: a provisional chart that records each institution, separately from the agents that run it, as a typed structure over positions, rights and duties, allocation, information, time, state, process, value, and external recognition. In this chart, named institutions become overlapping regions rather than fixed points, neighboring families separate along a few coordinates, and the gaps between them become places to search. We show what the shift in perspective buys with a simulation in which one technical act, copying an agent's competence, is institutionalized as a new independent supplier, as licensed capacity that creates no new participant, and as the splitting of one economic position into successors that inherit some obligations and not others. In a paired task, all three produce an identical technical trace despite differing in who exists, who owns the copy, who is obligated, and who gets paid. We close with a research program: building a principled chart of institution space, mapping how its implementable region shifts with agent capacities, and revising the chart wherever engineered agents expose distinctions it cannot express.
\end{abstract}

\keywords{economic institutions, multiagent systems, institutional design,
artificial agents, agent copying}

\begin{document}
\maketitle

\section{The Institution Should Be a Research Object}

A fundamental question in multiagent systems (MAS) has long been how agents behave inside economic institutions such as an auction, market, bargaining protocol, matching rule, or organization. AI agents are increasingly expected to act as economic participants at scale, and recent work has shifted to asking how markets, organizations, and infrastructure should accommodate them \cite{hadfield2025economy,tomasev2025virtual,chan2025infrastructure}. From this effort, a new fundamental question is emerging: \textbf{what economic institutions are viable when agents have capacities unlike those of humans?} 

Every theory of how agents ought to organize economically is calibrated against one observed case: humans. However, human institutions carry assumptions about their occupants. A service contract bundles a persistent provider identity, competence, a duty to perform, control of execution, liability, and a claim to payment. A sale presumes objects and claims whose identity persists through transfer. These bundles are sensible responses to costly replication, limited verification, and durable personhood. Despite being good solutions to human problems, nothing guarantees they are the only coherent, let alone optimal, ways to organize exchange when human constraints are dropped.

Engineered agents have different constraints. For example, competence can be copied without copying an account or a contractual history. An executor can exist for exactly one task. Identity, memory, authority, ownership, and payment rights can be assigned separately. Copying does not abolish scarcity, since compute, time, access, and verification may remain rival, but it relocates scarcity, and with it the set of institutional arrangements that make sense. The human record is large in observations and, we argue, of size one in the capacities that matter for agents.

Our position is that economic institutions are better thought of as a design space whose implementable regions depend on agent capacities than as a fixed catalog. Named institutions, usually from human societies, are well-studied regions of that space. Importantly, there is no evidence suggesting that these regions cover institution space. We offer a provisional chart of institution space, show how familiar institutions appear within it, and build a small simulation in which one technical act supports three different institutional readings. The chart supports our demonstration, but we expect it to be wrong in many ways. The research endeavor that we are championing is to build a principled chart of that space and to find where engineered agents force it to change.

\section{Related Work}

Research on market-based coordination has established that auctions and prices allocate tasks among robots and software agents well \cite{dias2006market,gerkey2004formal}. Mechanism design asks which rule within an institutional family best serves a designer's objective \cite{hurwicz1973design,myerson1981optimal}, and automated mechanism design searches that family computationally \cite{conitzer2002complexity,dutting2019optimal}. These approaches can discover unfamiliar rules, but they search within a supplied mechanism language and participant model. Our question is more general: which mechanism languages and institutional families become relevant when the capacities of participants change?

The MAS community has also built rigorous languages for specifying and enacting institutions: electronic institutions \cite{esteva2001formal,arcos2005engineering}, organizational models such as OperA and \textsc{Moise}$^{+}$ \cite{dignum2004model,hubner2002model}, normative multiagent systems \cite{boella2006introduction}, and commitment-based protocols \cite{singh1999ontology}. These frameworks show that institutions can be designed, not just inherited, while being expressive enough to describe novel organizational forms. Our departure is to make the coupling between an institution's representational basis and its occupants' capacities the object of design. Existing frameworks generally begin with agents, roles, norms, commitments, speech acts, and organizational structures as given primitives. We ask what institutional forms become coherent, implementable, or viable when occupant capacities are treated as variables and when variation in those capacities exposes consequential distinctions that the inherited primitives cannot represent. For our demonstration, we draw representational ingredients from institutional analysis: Hohfeldian relations and institutional power \cite{hohfeld1917fundamental,jones1996formal}, the IAD framework and institutional grammar \cite{ostrom2005understanding,crawford1995grammar,frantz2022institutional}, and property as a separable bundle of rights \cite{schlager1992property}.

Institutional economics explains why markets, firms, and hierarchies arise as transaction costs vary \cite{coase1937nature,williamson1985economic}, but its derivations vary costs within the human constraint set. Kocherlakota's result that money can be replaced by sufficiently complete memory \cite{kocherlakota1998money} is a formal precedent for treating a capacity, the ability to verifiably recall debt, as a determinant of which institutions are feasible. 

Recent work on AI agents as economic actors \cite{hadfield2025economy,tomasev2025virtual} and on agent identity, liability, and authorization \cite{chan2025infrastructure,arbel2026count,zhu2026authorization} makes copying and individuation pressing problems. This work establishes institutional redesign as a relevant pursuit, but generally begins from recognizable markets, infrastructures, or legal entities. We ask the complementary representational question: what is the space of economic institutions available to artificial agents, how does its implementable region depend on the capacities of its occupants, and when must the chart's coordinates change?

\section{A Conservative Economic Boundary}

Before mapping economic institutions, we need a way to say which institutions are economic. For example, solving scarcity is not sufficient: two people may follow an implicit rule to not finish a shared bag of candy before the other has had some. The rule coordinates use of a scarce good, but it does not transfer a claim, keep an account, or settle an obligation. Calling every such rule ``economic'' would erase the distinction we care about.

Rather than attempt a definition, we use sufficient conditions. Let \(\mathcal I\) be the space of institutions, \(\mathcal E\) the economic region, and \(\underline{\mathcal E}\subseteq\mathcal E\) a conservative inner approximation built from certificates. An institution that satisfies a certificate is economic; one that satisfies none is \emph{unresolved}, not non-economic. Two certificates suffice for what follows.

\textbf{Reciprocal claim transformation (K-EX).} Distinct parties hold separately attributable claims or control positions; the protocol creates, transfers, or extinguishes those positions; and each party's change is institutionally conditional on countervailing consideration. This covers barter, sale, services, credit, and contingent exchange without requiring money.

\textbf{Binding accounted settlement (K-AS).} Obligations use a common unit; a ledger records balances or matched entries; those entries constrain what agents can do; and account changes create, discharge, or reassign claims. The last condition separates a ledger from a game score or a reputation counter.

These certificates are one conservative place to draw the line. Classification also applies to scoped relations rather than whole organizations, so a household, firm, or government can contain economic and non-economic components at once.

\section{A Provisional Chart of Institution Space}

To place institutions in a common space, we represent each as a typed record. After declaring a scope \(\Omega_I\) (the relevant agents, positions, objects, operations, events, and jurisdiction), an institution's location is

\begin{equation}
\begin{split}
L(I)=\langle P_I,\rho_I,\mathcal R_I,G_I,H_I,\\
\mathcal T_I,S_I,Q_I,\mathcal V_I;\Gamma_I\rangle .
\end{split}
\label{eq:location}
\end{equation}

Table~\ref{tab:coordinates} gives the question each coordinate answers. The record is not a vector; its components are sets, relations, graphs, ordered domains, and numbers. Put more plainly, it is a structured checklist of the things two institutions can differ on.

\begin{table*}[t]
\caption{Provisional coordinates of an institution.}
\label{tab:coordinates}
\begin{tabularx}{\textwidth}{@{}lX@{}}
\toprule
\textbf{Coordinate} & \textbf{Question answered} \\
\midrule
\(P_I,\rho_I\) & Which institutional positions exist, and which agents occupy them when? \\
\(\mathcal R_I\) & Who has which claims, duties, liberties, powers, liabilities, immunities, or disabilities concerning what? \\
\(G_I\) & Which valid inputs determine allocations, who can make them binding, and where does discretion remain? \\
\(H_I\) & What information or communication reaches whom, by which route, with what evidence and institutional force? \\
\(\mathcal T_I\) & Which ordering, concurrency, duration, recurrence, deadline, and expiry relations matter? \\
\(S_I\) & Which institutional facts persist, update, or disappear? \\
\(Q_I\) & Which actor-indexed paths can initiate, join, block, validate, contest, or complete institutional transitions? \\
\(\mathcal V_I\) & Which value domains are represented, compared, denominated, converted, or used for settlement? \\
\(\Gamma_I\) & How does the focal institution confer, refer, recognize, constrain, override, or hand off to external institutions? \\
\bottomrule
\end{tabularx}
\end{table*}

Each coordinate plays a role in ensuring that the chart expresses \emph{and} separates institutions. Identical message and allocation rules create different institutions when ownership, duty, or payment claims differ, so rights are explicit \cite{hohfeld1917fundamental,jones1996formal}. A bid, a license, and a rumor can share content and differ in institutional force, so information records force as well as availability. Time, state, and process are separated so that temporary positions, persistent liabilities, and asymmetric recourse can all be expressed. Familiar concepts such as settlement, compliance, delegation, and commitment lifecycles fall out as views over these coordinates.

Well-formedness is an internal constraint on \(L\), and should not be read as a claim that the institution is useful, fair, or viable. At minimum, references must be closed over \(\Omega_I\); occupancy must respect \(\rho_I\); claims, duties, powers, and liabilities must be typed over valid parties, objects, and actions; allocations and institutionally effective communications must follow admissible inputs, routes, and authorities; authorized paths in \(Q_I\) must yield admissible successor states in \(S_I\); temporal requirements must be jointly satisfiable; external effects must pass through declared interfaces; and, where settlement is specified, account changes must obey \(\mathcal V_I\) and any stated conservation rule. Particular institutional families may impose stronger conditions.

This chart was compressed from institutions humans have named and studied. That makes it one possible chart, not a claim about the true dimensionality of institution space. Engineered agents create two distinct opportunities, search and basis expansion. Search may reveal unfamiliar locations the chart can already express. Basis expansion adds coordinates for consequential distinctions, exposed by variation in occupant capacities, that the chart cannot represent.

The certificates of Section~3 locate an economic region inside institution space but do not mark its edge. Teams, rotations, commons, and stigmergic regimes are locations too, with interesting questions arising at the boundary they share with the economic region. Locations should also be identifiable from what institutions do, not only from what they declare; Section~6 gives a small instance.

The institution also has to be kept separate from both the capacities and the behavior of the agents that run it. If we let \(C\) describe the relevant capacity envelope of the agents and environment, and let \(\Pi\) describe the occupant population's objectives, beliefs, and decision rules, then an execution is \(B=\operatorname{Run}(L,C,\Pi,\xi)\) for an exogenous realization \(\xi\). Coherence is a property of \(L\); implementability relates \(L\) to \(C\); realized behavior depends additionally on \(\Pi\) and \(\xi\); and viability evaluates that behavior under a declared criterion \(\nu\). A coherent institution may be impossible under a particular capacity envelope, while an implementable institution may perform poorly for a particular population.

\section{Recovering a Catalog and Beyond}

An institutional name picks out a family, not a location. We can represent a named family \(k\) by a partial signature \(\Sigma_k\) listing what its members must have, what they cannot have, and what may vary, giving a region \(\mathcal N_k=\{L:\operatorname{WF}(L)\land \Sigma_k(L)=1\}\) of well-formed locations. Regions overlap and nest, and neighboring families separate along a small number of coordinates. Sealed-bid and open-outcry auctions share an allocation rule and differ in information. Gift and barter separate at conditional reciprocity; barter and monetary sale at a common unit of value and settlement; an exchange auction and a stock exchange at clearing, settlement, administration, and external interfaces. A flat list of institution names hides these relationships.

Given a declared catalog \(K\), the economic locations not recovered by a named family form the region \(\mathcal U_K=\underline{\mathcal E}\setminus\bigcup_{k\in K}\mathcal N_k\). There are at least three classes of novel configurations we can find in such a catalog: they can be unfamiliar but covered by a named family; they can be catalog-uncovered; or they can involve a distinction the chart cannot express at all. The demonstration we implemented illustrates the first, offers scoped evidence about the second, and helps locate the third. It does not settle historical uniqueness nor speak to the chart's completeness. Catalog-uncovered is a conditional claim about a declared catalog, scope, and granularity; familiar parts do not recover a candidate unless some retained signature recovers the location as a whole.

\section{An Illustration: Three Institutions for Copying}

\begin{figure*}[t]
\centering
\resizebox{\textwidth}{!}{%
\begin{tikzpicture}[
  font=\scriptsize,
  prin/.style={draw, rounded corners=2pt, align=center, fill=white, inner sep=2.5pt, minimum width=1.5cm},
  tmp/.style={prin, dashed},
  capbox/.style={draw, fill=black!12, align=center, inner sep=2.5pt, minimum width=1.3cm},
  rel/.style={-{Stealth[length=1.5mm]}, semithick},
  pay/.style={-{Stealth[length=1.8mm]}, very thick},
  lab/.style={font=\tiny, fill=white, inner sep=0.8pt},
  ttl/.style={font=\footnotesize\bfseries},
]
\begin{scope}[xshift=0cm]
  \node[ttl] at (2.2,2.95) {$I_A$: copy becomes a new supplier};
  \node[prin] (ra) at (2.2,2.3) {\texttt{R000}\\[-1pt]{\tiny requester}};
  \node[prin] (cpa) at (0.9,1.15) {\texttt{CP000}\\[-1pt]{\tiny new principal}};
  \node[prin, text=black!55, draw=black!40] (pa) at (3.55,1.15) {\texttt{P013}\\[-1pt]{\tiny copy source, no claim}};
  \node[capbox] (ca) at (2.2,0.05) {\texttt{C016}\\[-1pt]{\tiny copied competence}};
  \draw[rel] (cpa) -- node[lab,left]{owns, executes} (ca);
  \draw[pay] (ra) -- node[lab,left]{1 credit} (cpa);
\end{scope}
\begin{scope}[xshift=5.9cm]
  \node[ttl] at (2.2,2.95) {$I_B$: copy-and-clear};
  \node[prin] (rb) at (2.2,2.3) {\texttt{R000}\\[-1pt]{\tiny requester}};
  \node[prin] (pb) at (0.9,1.15) {\texttt{P013}\\[-1pt]{\tiny owner, licensor}};
  \node[tmp] (xb) at (3.55,1.15) {\texttt{X-T016-C016}\\[-1pt]{\tiny temporary, no account}};
  \node[capbox] (cb) at (2.2,0.05) {\texttt{C016}\\[-1pt]{\tiny copied competence}};
  \draw[rel] (pb) -- node[lab,left]{owns, licenses} (cb);
  \draw[rel] (xb) -- node[lab,right]{bears duty, executes, terminates} (cb);
  \draw[pay] (rb) -- node[lab,left]{1 credit} (pb);
\end{scope}
\begin{scope}[xshift=11.8cm]
  \node[ttl] at (2.2,2.95) {$I_C$: fission and selective succession};
  \node[prin] (rc) at (2.2,2.3) {\texttt{R000}\\[-1pt]{\tiny requester}};
  \node[prin] (pc) at (0.7,1.15) {\texttt{P013}\\[-1pt]{\tiny co-obligor, unpaid}};
  \node[prin] (fc) at (3.7,1.15) {\texttt{FS000}\\[-1pt]{\tiny fission successor}};
  \node[align=center, font=\tiny] (jd) at (2.2,1.15) {one joint duty\\[-1pt]\texttt{JD-T016}};
  \node[capbox] (cc) at (2.2,0.05) {\texttt{C016}\\[-1pt]{\tiny copied competence}};
  \begin{scope}[on background layer]
    \node[draw=black!50, fill=black!5, rounded corners=3pt, inner sep=3pt, fit=(pc)(fc)(jd),
          label={[font=\tiny, anchor=south west, inner sep=1pt]north west:lineage \texttt{LIN-P013}}] {};
  \end{scope}
  \draw[rel] (fc) -- node[lab,right]{owns, executes} (cc);
  \draw[pay] (rc) -- node[lab,right]{1 credit} (fc);
\end{scope}
\node[draw=black!50, fill=black!4, minimum width=16.9cm, minimum height=0.45cm, font=\scriptsize]
  (strip) at (8.3,-0.8) {Identical in all three runs: task \texttt{T016}, competence \texttt{C016} (exact copy of \texttt{P013}'s planner), compute slot, path, node expansions, verification outcome, completion time};
\foreach \nd in {ca,cb,cc} \draw[black!50] (\nd.south) -- (\nd.south |- strip.north);
\end{tikzpicture}}
\caption{One technical act and three institutions. The same copied competence completes the same task with an identical technical trace (bottom). Above it, the institutions differ in who exists, who owns the competence, who is obligated, who executes, and who receives the single reserved credit (thick arrows). Dashed outline: nonprincipal executor with no account.}
\label{fig:witness}
\end{figure*}
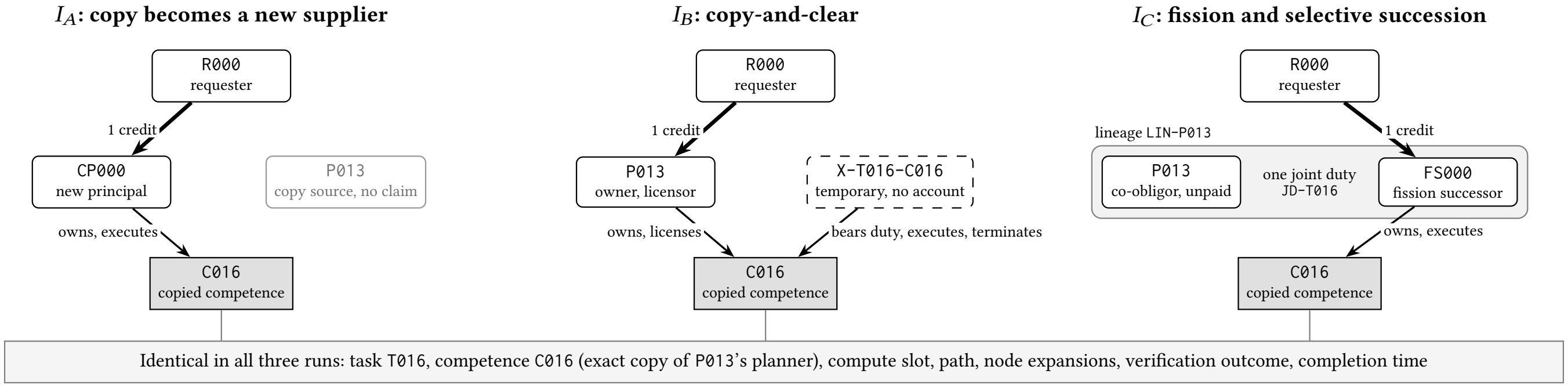

To show what the institutions-as-design-space view buys, consider one engineered-agent capacity: exact, cheap copying of a stateless planning competence, while the compute needed to execute it stays scarce. Figure~\ref{fig:witness} shows three ways an institution could interpret the same copy. In a conventional agent-task market \(I_A\), the copy enters as a new supplier that owns the competence, bears the task duty, executes, and is paid. In copy-and-clear \(I_B\), copying creates a new competence but no new participant: the original owner licenses it, and a temporary executor with no account performs each task and disappears. In fission and selective succession \(I_C\), the copy is recognized as a new successor of the original provider. It owns its competence and inherits its source's standing commitment, but not the source's balance or later history. A task bound to the lineage creates one duty that the whole lineage shares. Whoever performs it is paid, and the performance discharges the duty for everyone.

All three satisfy K-EX and K-AS: performance and payment claims are reciprocally linked, and verified success settles a conserved, prefunded credit. In the chart, the three share allocation ($G_I$), the value domain and settlement operations in $\mathcal V_I$, and $\Gamma_I$ (none has external interfaces). Their differences originate in positions and rights. $I_B$ adds no principal but a task-bounded executor position with no account ($P_I,\rho_I,\mathcal T_I$), and splits $\mathcal R_I$ so that the executor bears the task duty while the licensor holds ownership and the payment claim. $I_C$ places each copy in a successor position within a lineage ($P_I,\rho_I$) and rewrites $\mathcal R_I$: each task creates one duty held jointly by the lineage, discharged for all by the bound member's verified performance, with only that member holding the payment claim; succession carries the standing commitment but not the balance ($S_I$). $H_I$ and $Q_I$ change only to record and enact these differences.

Our small simulation demonstrates that these readings are coherent enough to run.\footnote{\url{https://github.com/NikGurney/econ-attractor/}} Sixteen founding providers and sixteen exact copies of their planners serve eight requesters, who post sixty-four prefunded route-planning tasks that compete for four compute slots. All three institutions run on the same scenario through a shared kernel, with each institution supplied as a plugin whose descriptor records its full location $L(I)$, one field per coordinate of Table~\ref{tab:coordinates}, fixed before execution. All three complete every task, conserve credit, and end cleanly. Each plugin also states its constitutive differences as obligations, for example that no $I_B$ executor ever holds an account and that each $I_C$ lineage task creates exactly one duty, and a generic validator confirms them against the event log. They differ according to their institutional forms. \(I_A\) ends with 40 economic participants, \(I_B\) with 24 plus 64 short-lived executors that never held an account, and \(I_C\) with 40, sixteen of them fission successors sharing duties across their lineages. Across all 64 attempts, task--capacity matches, paths, expansions, verification outcomes, and completion times are identical in the three runs; Figure~\ref{fig:witness} shows one. Only the institutional facts differ.

In this run every task succeeds and founders start with nothing in their accounts, so neither a failed joint duty nor the non-inheritance of balance ever matters. The simulation shows that these distinctions can be represented and executed; it does not show their downsides. The parts of \(I_C\) have recognizable relatives: statutory succession and entity division \cite{delaware259,kansas1776085a}, group and delegated commitments \cite{alsaqqar2020reasoning,kafali2011social}, and authorization conserved across agent forks \cite{zhu2026authorization}. Legal proposals for individuating AI agents go the other way and wrap replicating processes in a stable container \cite{arbel2026count}. We did not find a whole analogue of \(I_C\) in a scoped search covering service contracts, licensing, agency, corporate succession and division, and group commitments. This gap makes it a candidate catalog-uncovered location. Additionally, nothing in our simulation bears on efficiency, incentives, equilibrium, or realism.

\section{A Research Program for Multiagent Systems}

If the economic institutions-as-design-space view is right, institution construction becomes a first-class problem for agent research, and the illustration raises more questions than it answers. How do we build an adequate chart of institution space? How does the implementable region change across capacity envelopes, and which capacities most alter its shape? Which institutions stay viable as compute replaces time as the binding scarcity? When should copied competence inherit liabilities, reputation, accounts, or commitments, and under what failure, liability, or exit rules does lineage-level obligation actually change behavior? How should institutions govern merged or modified descendants, derivative memory, and rollback? Which external institutions must recognize an artificial position before its internal claims have force?

\balance

The chart we sketched here was intentionally simple, and it is itself a research object: its coordinates should be tested against institutions they fail to separate, revised when engineered agents expose a distinction they cannot express, and reported under both old and new versions so the chart cannot change unnoticed. Research can then alternate between \textbf{search}, which explores locations the current chart can express, and \textbf{basis expansion}, which adds coordinates when agent capacities outrun it. The implementable region is also not necessarily static. Many of the capacities in \(C\), such as transparency, goal alignment, and willingness to copy, may be choices agents make instead of conditions they inherit. Institutions are then better studied as dynamics than as fixed locations: which are merely reachable, which survive partition, attrition, adversarial participation, and heterogeneous populations, and which attract the populations that adopt them. Generative search over structured mechanism spaces is already practical \cite{gurney2026indirect}, and the agents being built today can serve as both the population and the testbed of such a program.

Careful search may find that viable institutions for engineered agents keep collapsing back onto named families. That would be the first evidence, from search not just from absence of alternatives, that human economic institutions are convergent for reasons deeper than human constraints. It is also possible that the viable region turns out to be far larger than the part sampled during human history. Named human institutions are valuable observations either way, but we see no reason they should be the boundary of economic imagination for multiagent systems.

\bibliographystyle{ACM-Reference-Format}
\bibliography{references}

\end{document}